\documentstyle[sprocl,psfig,epsfig]{article}

\def\be{\begin{equation}}
\def\ee{\end{equation}}
\def\bea{\begin{eqnarray}}
\def\eea{\end{eqnarray}}
\newcommand{\vs}{\vspace{-0.005cm}}

\begin{document}

\hfill {{\small FZJ-IKP(TH)-1999-09}}

\smallskip

\title{ISOSPIN VIOLATION IN THE TWO--NUCLEON SYSTEM}

\author{E. Epelbaum, \underline{Ulf-G. Mei{\ss}ner}}

\address{FZ J\"ulich, IKP(TH), 
D-52425 J\"ulich, Germany\\E-mail: Ulf-G.Meissner@fz-juelich.de}

\maketitle\abstracts{I discuss isospin violation in an effective
field theory appraoch of the nucleon--nucleon interaction. The observed
leading order charge independence breaking  is explained
in terms of one--pion exchange together with a four--nucleon contact
term. A general classification scheme of corrections to this leading
order results is presented and an explanation is offered why 
charge--dependent pion--nucleon coupling constants can not play a role in the
observed isospin violation in the NN scattering lengths. Charge
symmetry breaking, the size of $\pi\gamma$--exchange graphs
 and the pionless theory are also discussed.}

\section{Introduction}

Heisenberg introduced the notion of isospin in 1930 as a label to distinguish 
between neutrons and protons when he was studying atomic nuclei. 
He noticed that the clustering of stable nuclei around $N \simeq Z$
implied a new (internal) symmetry. This was later extended to the
pion triplet, which is believed to mediate the long range part of the
nuclear force. Heisenberg's ideas were used by Breit and collaborators (and others)
to establish the so--called charge independence, observing that after
removal of the Coulomb effects between protons, the $pp$, $np$ and $nn$
force were equal to a good precision. A weaker condition is the so--called
charge symmetry, which states that the $pp$ and $nn$ force should be
equal with no assertion to the $np$ force. Since that time, QCD has
emerged as the theory of the strong interactions. In the absence of 
electroweak interactions, isospin symmetry is observed in the sector
of the two lightest quark flavors for {\it equal} masses, $m_u = m_d$,
i.e. the QCD Hamiltonian is invariant under a global (flavor) transformation 
of the type
\be
q  = 
\left( \begin{array}{c} 
                 u \\ d
\end{array} \right) 
\rightarrow q' = {\cal V}\, q = {\cal V} \,
\left( \begin{array}{c} 
                 u \\ d
\end{array} \right) \,\,\, ,
\quad {\cal V} \in SU(2)
\quad .
\label{ud}
\ee
Of course, in nature the light quark masses are not equal. In fact, they
differ considerably, $m_u/m_d = 0.55$. Still, isospin is a good approximate
symmetry because what counts is not the ratio of the quark masses but rather
their difference compared to the typical scale of strong interactions, 
$(m_d-m_u)/\Lambda_{\rm QCD} \ll 1$. QCD thus provides a reason why
isospin is such a good approximate symmetry. This can hardly be
considered ``accidental''. In the presence of electromagnetism, further isospin
violation is induced since the charges of the quarks are unequal. These effects are
in general small due to the explicit apperance of the fine structure
constant $\alpha \simeq 1/137$. In fact, in many cases the strong and
electromagnetic isospin--violating effects are of the same size. A good
example is the neutron--proton mass difference. On the quark level, charge 
symmetry is a rotation about the 2--axis in isospin space
\be
{\cal P}_{\rm cs}
\left( \begin{array}{c} 
                 u \\ d
\end{array} \right) 
= {\rm e}^{(i\pi /2)\tau_2} \,
\left( \begin{array}{c} 
                 u \\ d
\end{array} \right) 
=
\left( \begin{array}{c} 
                 -d \\ \,\,u
\end{array} \right) \,\,\, ,
\quad {\cal P}_{\rm cs}^2 = {\bf 1}
\quad .
\label{Pcs}
\ee
Our aim is get some insight about isospin symmetry (charge independence)
and charge symmetry in the two--nucleon system based on an effective
field theory approach. On the level of the two--nucleon force,
charge independence implies a potential of the form 
\be
V_{ij} = A + B \tau_{(i)} \cdot \tau_{(j)}~,
\ee
where $A$ and $B$ are functions of the nucleon spin and momentum (or space)
coordinates and `$(i),(j)$' labels the interacting nucleons. 
Charge symmetry allows for a more general two--body force,
\be
V_{ij, {\rm cs}} = A + B \tau_{(i)} \cdot \tau_{(j)} + C \tau_{(i)}^3
\cdot \tau_{(j)}^3~.
\ee
If one works out the Coulomb potential in terms of the nucleon charge
matrix $Q = e (1+\tau^3)/2$, it is obvious the electromagnetic effects lead
to breaking of charge independence and charge symmetry,
\be
V_{ij, {\rm Coulomb}} = {e^2\over 4r^2} (1+ \tau_{(i)}^3)
(1+ \tau_{(j)}^3)~.
\ee 
I now wish to show how the question of
isospin violation can be addressed in an effective field
theory approach to the NN interactions. There are two equivalent ways to
proceed. One is to follow the suggestion of Weinberg~\cite{wein} and perform
the chiral expansion on the level of the potential (i.e. for the irreducible diagrams)
and iterate this potential in the Lippmann--Schwinger equation to generate 
the bound and scattering states.
The other is to directly expand the amplitude after performing a nonperturbative
resummation for the S--waves.\cite{KSW} The first path has been followed by
van Kolck~\cite{bira} and I will consider the second option here.\cite{em}
Pions are included and treated perturbatively in this approach. For
the topic addressed here, the discussion whether or not pions should
be treated in such a way is inessential. In fact, as shown in
section~6, one can even work with a pionless theory. In that case,
however, the physical interpretation of CIB and CSB becomes much less
transparent.

\section{Facts, explanations and  mysteries}

It is well established that the nucleon--nucleon interactions
are charge dependent (for a review, see e.g.\cite{jerry} and a recent
summary is given in~\cite{SAC}). For example, in 
the $^1S_0$ channel one has for the scattering lengths $a$ and the effective
ranges $r$ 
\bea\label{CIBval}
\Delta a_{\rm CIB} &=& \frac{1}{2} \left( a_{nn} + a_{pp} \right) - a_{np}
= 5.7 \pm 0.3~{\rm fm}~,\\
\Delta r_{\rm CIB} &=& \frac{1}{2} \left( r_{nn} + r_{pp} \right) - r_{np}
= 0.05 \pm 0.08~{\rm fm}~.
\eea
These numbers for charge independence breaking ({\bf CIB})
are based on the Nijmegen potential~\cite{nij} and the Coulomb effects
for $pp$ ($nn$) scattering are subtracted based on standard methods (for a 
treatment of this in an EFT framework see refs.\cite{RK,BRH}). 
The charge independence breaking in the scattering lengths is large, of
the order of 25\%, since $a_{np} = (-23.714 \pm 0.013)\,$fm. Of course,
it is magnified at threshold due to kinematic factors (as witnessed
by the disapperance of the effect in the effective range).
In addition, there are
charge symmetry breaking ({\bf CSB}) effects leading to different values for
the $pp$ and $nn$ threshold parameters,
\be\label{CSBval}
\Delta a_{\rm CSB} =  a_{pp} - a_{nn} 
= 1.5 \pm 0.5~{\rm fm}~,\quad
\Delta r_{\rm CSB} =  r_{pp} - r_{nn} 
= 0.10 \pm 0.12~{\rm fm}~.
\ee
Combining these numbers gives as central values $a_{nn} = -18.8\,$fm
and $a_{pp} = -17.3\,$fm. Both the CIB and CSB effects have been 
studied intensively within potential
models of the nucleon--nucleon (NN) interactions. In such approaches, the
dominant CIB comes from the charged to neutral pion mass difference in the
one--pion exchange (OPE), $\Delta a_{\rm CIB}^{\rm OPE} \sim 3.6 \pm 0.2\,$fm. 
Additional contributions come from $\gamma\pi$ and $2\pi$ (TPE) exchanges. 
According to some calculations, their size is approximately $1/3^{\rm rd}$ 
of the leading OPE contribution. For the TPE, significantly smaller results
can also be found in the literature, see e.g. table~3.3 in ref.\cite{jerry}
In case of the $\pi\gamma$ contribution, a recent calculation\cite{CF} gives
a value which is more than a factor of ten smaller than 
the leading OPE contribution. 
We will come back to these issues later (see section~5).
Note also that the charge dependence in the pion--nucleon coupling constants 
(if existing) in OPE and TPE
almost entirely cancel. Naively, one would expect a possible charge dependence
of the pion--nucleon coupling constants to be of a similar importance as the
pion mass difference in OPE. This strong suppression of
charge--dependent couplings has
so far been eluded a deeper understanding. We will come back to this later
on. In the meson--exchange picture, CSB originates mostly from  
$\rho-\omega$ mixing, $\Delta a_{\rm CSB}^{\rho-\omega} \sim 1.2 \pm 0.4\,$fm. Other
contributions due to $\pi-\eta$, $\pi-\eta'$ mixing or the proton--neutron
mass difference are known to be much smaller. It will be shown in what follows
that EFT gives novel insight concerning the size of the TPE and $\pi\gamma$
contributions as well as the suppression of possible charge--dependent $\pi N$
coupling constants.

\section{Effective Lagrangian and power counting}

The effective
Lagrangian underlying the analysis of isospin violation in the two--nucleon
system consists of a string of terms,
\be
{\cal L}_{\rm eff} = {\cal L}_{\pi\pi} + {\cal L}_{\pi N} +{\cal L}_{NN}~.  
\ee
Each of these terms splits into a strong and an electromagnetic one.
The strong contributions include (besides many others) the 
isospin--violation due to the quark mass difference
$m_u - m_d$ whereas the em terms stem from the
dynamics of the virtual photons (photon loops and em counterterms).
To include virtual photons in the pion and the pion--nucleon
system is by now a standard procedure.\cite{urech}$^-$\cite{ms} 
We note that in the pion
and pion--nucleon sector, one can effectively count the electric charge as
a small momentum or meson mass. This is based on the observation that
$M_\pi / \Lambda_\chi \sim e/\sqrt{4\pi} = \sqrt{\alpha} \sim 1/10$ 
since $\Lambda_\chi \simeq 4\pi f_\pi = 1.2\,$GeV. It is thus possible to 
turn the dual expansion in small momenta/meson masses on one side and in 
the electric coupling
$e$ on the other side into an expansion with {\it one} generic small parameter.
We also remark that from here we use the fine structure constant $\alpha
=e^2/4\pi$ as the em expansion parameter. Because of the additional
$1/4\pi$ factor (from integrating out hard photons) one expects the
corresponding em constants  to be of order one (of natural size). A
dimensional analysis of such terms in the pion--nucleon sector is given
in ref.\cite{mm} and similarly for the pion sector in 
refs.\cite{urech}$^,$\cite{marc}.

We now turn to the two--nucleon sector. One can either expand the
potential (irreducible diagrams) or directly the amplitudes.
A complication arises due to the unnaturally large S--wave scattering
lengths, $1/a \ll M_\pi$.  A {\it systematic} power counting scheme to deal 
with this directly on the level of the scattering amplitude
is  the recently proposed power divergence 
subtraction method (PDS) of Kaplan, Savage and Wise 
(KSW)~\cite{KSW}.\footnote{There exist by now modifications
of this approach and it as been argued that it is equivalent to cut--off
schemes. We do not want to enter this discussion here but rather stick
its original version.} Essentially, one resums the lowest
order local four--nucleon contact terms $\sim C_0 (N^\dagger N)^2$ (in
the S--waves) to generate the large scattering lengths and treats the remaining
effects perturbatively, in particular also pion exchange. This means that
most low--energy observables are dominated by contact interactions. 
The chiral expansion for NN scattering entails a new scale $\Lambda_{NN}$ of the
order of 300~MeV, so that one can systematically treat external
momenta up to the size of the pion mass. There have been suggestions
that the radius of convergence can be somewhat enlarged~\cite{mestew},
but in any case $\Lambda_{NN}$ is considerably smaller than the typical
scale of about 1~GeV appearing in the pion--nucleon sector. 
It is straigthforward to extend the underlying effective Lagrangian to include
the strong isospin--violating and electromagnetic  four--fermion contact
interactions. Consider first the strong terms. 
Up to one quark mass insertion, the effective Lagrangian takes the form (note that
terms involving the Pauli isospin matrices $\vec{\tau}$ can be
obtained by Fierz reordering)
\bea
{\cal L}_{NN}^{\rm str}&=& l_1 (N^\dagger N)^2 + l_2 (N^\dagger \vec{\sigma} N)^2
+ l_3 ( N^\dagger \langle \chi_+ \rangle N) (N^\dagger N) + l_4
(N^\dagger \hat{\chi}_+ N)(N^\dagger N)
\nonumber \\
&+& l_5 ( N^\dagger \vec{\sigma}\langle \chi_+ \rangle N) (N^\dagger \vec{\sigma}
 N) + l_6 (N^\dagger \vec{\sigma} \hat{\chi}_+ N)(N^\dagger \vec{\sigma} N)
+ \ldots~,
\eea
where $\langle \ldots \rangle$ denotes the trace in flavor space and
$\chi_+$ contains the light quark masses. $\hat\chi_+ = \chi_+ - \langle \chi_+
\rangle /2$ is only non--vanishing for $m_u \neq m_d$ and 
the ellipsis denotes terms with two (or more) derivatives acting on the
nucleon fields. Similarly, one can construct the em terms. The ones without
derivatives on the nucleon fields  read (possible terms involving the
Pauli isospin matrices are of no relevance for the following)
\bea
{\cal L}_{NN}^{\rm em}&=& N^\dagger \left\{ r_1 \langle Q_+^2 - Q_-^2\rangle
+ r_2 \hat{Q}_+ \langle Q_+\rangle \right\} N (N^\dagger N) 
\nonumber \\
&+& N^\dagger \vec{\sigma} \left\{ r_3 \langle Q_+^2 - Q_-^2\rangle 
+ r_4  \hat{Q}_+ \langle Q_+\rangle \right\} N (N^\dagger \vec{\sigma} N)
\nonumber \\
&+&   N^\dagger  \left\{ r_{5}Q_+ + r_{6}\langle Q_+ \rangle \right\} 
N (N^\dagger Q_+ N) \nonumber \\ &+&
   N^\dagger \vec{\sigma} \left\{ r_{7}Q_+ + r_{8}\langle Q_+ \rangle 
\right\} N (N^\dagger \vec{\sigma} Q_+ N) \nonumber \\
&+& r_{9} (N^\dagger  Q_+ N)^2 + r_{10} (N^\dagger \vec{\sigma} Q_+ N)^2~,
\eea
with $Q_\pm = uQ^\dagger u \pm u^\dagger Q u^\dagger$  and $Q$ the nucleon
charge matrix. All these terms are due to the integration of the
``hard'' virtual photons. The ``soft'' ones are contained in the
pertinent covariant derivatives acting on the pion or nucleon fields,
see e.g. refs.\cite{mms}$^,$\cite{ms}.
We remark that since $\Lambda_{NN}$ is significantly smaller than $\Lambda_\chi$,
it does not pay to treat the expansion in the generic KSW momentum\footnote{So 
that no confusion with the charge matrix can arise, we denote this
parameter with an extra ``tilde'' contrary to common use. A pedagogical
discussion of this expansion parameter can be found in ref.\cite{KaBonn}.} 
\be
{\tilde{Q} / \Lambda_{NN}} \sim {\mu / \Lambda_{NN}}
\sim {M_\pi / \Lambda_{NN}}
\ee
(with $\mu$ the scale of dimensional regularization)
simultaneously with the one in the fine structure 
constant (as it is done e.g. in the pion--nucleon sector). Instead, one has
to assign to each term a {\it double expansion parameter},
\be 
(\tilde{Q}/\Lambda_{NN})^n \, \alpha^{m}~, \quad 
\tilde{Q}/\Lambda_{NN}  \simeq 1/3~, \quad \alpha = 1/137~,
\ee
with $n$ and $m$ integers ($m \ge 1$ and $n$ can be negative). 
Clearly, even if we work at 
next--to--next--to leading order (NNLO) which entails a supression
by two powers in $\tilde{Q}$, we need only be concerned with the
leading em terms $\sim \alpha$ since $\alpha \ll 1$.
Lowest order charge independence breaking is due to a term $\sim (N^\dagger
\tau^3 N)^2$ whereas charge symmetry breaking at that order is given
by a structure $\sim (N^\dagger \tau^3 N) (N^\dagger N)$.
In the KSW approach, it is customary to project the Lagrangian terms on the
pertinent NN partial waves. Denoting by $\beta$ the  $^1S_0$ partial
wave for a given cms energy $E_{\rm cms}$, the Born amplitudes for the lowest
order CIB and CSB operators between the various two--nucleon states
takes the form
\bea\label{LNNME}
\langle \beta , pp | {\cal L}^{\rm em}_{NN} | \beta , pp\rangle &=&
-\left(\frac{Mp}{2\pi}\right)  \alpha 
\left( E_0^{(1)} (\mu ) +   E_0^{(2)} (\mu ) \right)~,
\nonumber\\
\langle \beta , nn | {\cal L}^{\rm em}_{NN} | \beta , nn\rangle &=&
-\left(\frac{Mp}{2\pi}\right)  \alpha 
\left( E_0^{(1)} (\mu ) -   E_0^{(2)} (\mu ) \right)~,
\eea
where the coupling constant $E_0^{(1,2)} (\mu )$  will be determined later on.
{}From the renormalization group equations (RGE) it will  also be possible to
derive the scaling properties of the $E_{2n}^{(1,2)} (\mu )$ for four--nucleon
contact operators with $2n$ derivatives.  The terms with the 
superscript '$(1)$' refer to CIB whereas the second ones
relevant for CSB  are denoted by the superscript '$(2)$'. Higher order operators
are denoted accordingly. There is, of course, also a CIB contribution to the
$np$ matrix element. To be consistent with the charge symmetric calculation
of ref.\cite{KSW}, one has to  absorb its effect in the constant $D_2$, i.e. it
amounts to a finite renormalization of $D_2$ and is thus not observable.
In eq.(\ref{LNNME}), $p= \sqrt{ME_{\rm cms}}$ is the nucleon cms momentum and
$M$ the nucleon mass.

\section{Isospin violation in nucleon--nucleon scattering}

Let me now work out the pertinent $^1S_0$ phase shifts.
The $np$ case has already been calculated in ref.\cite{KSW}, the
leading amplitude being of order $\tilde{Q}^{-1}$. 
We want to calculate the leading em corrections (here, ``Cs'' means
Coulomb--subtracted and we neglect the small magnetic corrections for the 
$nn$ system),
\be
\Delta {\cal A} = {\cal A}_{nn} - {\cal A}_{np}
= {\cal A}_{pp}^{\rm Cs} - {\cal A}_{np}~.
\ee
The pertinent diagrams are shown in fig.~1.
Besides the contact terms written down in eq.(\ref{LNNME}), we have to
consider the pion mass difference in one--pion exchange.
OPE diagrams with different pion masses
have the isospin structure $O_{12} = \tau_{(1)a} \Delta_\pi^{ab} \tau_{(1)b}
= (a+b)\, \vec{\tau}_{(1)} \,  \vec{\tau}_{(2)}
-b \, {\tau}_{(1)}^3 \, {\tau}_{(2)}^3$ and obviously lead to CIB.
{}From the first graph in fig.~1 one easily concludes that
these effects are of order $\alpha \tilde{Q}^{-2}$ since the diagram
contains two derivative $\pi N$ couplings of order $\tilde{Q}$ and two pion propagators
of order $\tilde{Q}^{-2}$. 
The notation follows closely the one of KSW in that the
corresponding three amplitudes are called ${\cal A}_{2,-2}^{II,III,IV}$.
Here, the first (second) subscript refers to the power in $\alpha$ 
($\tilde{Q}$) and the superscripts to the first three diagrams of the figure. 
This gives
\begin{center}
\begin{figure}[tb]
\centerline{\psfig{file=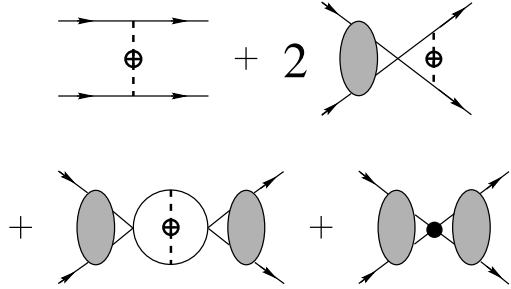,width=1.5in,angle=270}}
\caption{
Relevant graphs contributing to charge independence breaking at 
leading order $\alpha \tilde{Q}^{-2}$. The blob stands for the resummation
of the lowest order $(N^\dagger N)^2$ contact term $\sim C_0$. 
The open (filled)
circle denotes a pion mass insertion $\sim \delta m^2$ (an insertion
of the leading four--nucleon operators $\sim \alpha E_0^{(1)}$). For
charge symmetry breaking, the leading contribution is given by the
last diagram with the filled circle denoting an insertion
$\sim \alpha E_0^{(2)}$.
\label{fig:A}}
\end{figure}
\end{center}
\noindent
\bea
\Delta {\cal A}^{II}_{1,-2} &=&  \Gamma \left[ \frac{1}{4p^2} \ln
\left(1 + \frac{4p^2}{m^2}\right) - \frac{1}{m^2+4p^2} \right]~,
\nonumber\\ && \nonumber \\
\Delta {\cal A}^{III}_{1,-2} &=& \Gamma \, {\cal F} \,
  \left[ \frac{1}{pm} \arctan
  \frac{2p}{m} + \frac{i}{2pm}\ln \left( 1 + + \frac{4p^2}{m^2}\right)
  - \frac{1 + \frac{2ip}{m}}{m^2+4p^2} 
\right]\nonumber\\ && \nonumber \\
\Delta {\cal A}^{IV}_{1,-2} &=& \Gamma 
\, {\cal F}^2 \left[ \frac{i}{m^2} \arctan
  \frac{2p}{m} - \frac{1}{2m^2}\ln \left( \frac{m^2+4p^2}{\mu^2}\right)
  +\frac{1}{m^2} -\frac{1}{2} \frac{1 + \frac{2ip}{m}}{m^2+4p^2} 
\right]\nonumber\\ && \nonumber \\
\Gamma &=& -\delta m^2 \frac{g_A^2}{2f_\pi^2}~, 
\quad {\cal F} = \left( \frac{mM{\cal A}_{-1}}{4\pi}\right)~,
\quad \delta m^2 =
m_{\pi^\pm}^2 - m_{\pi^0}^2~, \\ && \nonumber
\eea
with ${\cal A}_{-1} \equiv {\cal A}_{0,-1}$ the 
leading term in the expansion of the $np$ $^1S_0$ amplitude,
\be\label{Amo}
{\cal A}_1 = {C_0 \over 1+  {C_0 M \over 4\pi} (\mu + ip ) }~.
\ee
Furthermore,
$f_\pi = 92.4\,$MeV is the pion decay constant, $g_A = 1.26$ the axial--vector
coupling constant and $m^2 = 2 m_{\pi^\pm}^2 - m_{\pi^0}^2$ the 
``average'' pion mass. The somewhat unusually looking expressions for
$m^2$ and $\delta m^2$ are derived in  appendix~A.    
Note that while the diagrams II and III are
convergent, IV diverges logarithmically. Therefore, the Lagrangian
must contain a counterterm of the structure 
$E_0^{(1)} (\mu ) \alpha (N^\dagger \tau^3 N)(N^\dagger \tau^3 N)$
to make the amplitude scale--independent. The leading CIB contribution
thus scales as $\tilde{Q}^{-2}$ and for higher order operators with $2n$ 
derivatives we can establish the scaling property $E_{2n}^{(1)} 
\sim \tilde{Q}^{-2+n}$. This does
not contradict the KSW power counting for the isospin symmetric theory
since $\alpha \ll 1$.  The insertion from this contact term is shown in the 
last diagram of  fig.~1 and leads to an
additional contribution to $\Delta {\cal A}$. In complete analogy,
we can treat the leading order CSB effect which is due to an
operator of the form $\alpha E_0^{(2)}(N^\dagger \tau_3 N)(N^\dagger N)$.
This term is, however, finite. Putting pieces together, this gives
for the $nn$ and $pp$ case 
\bea\label{AE}
\Delta {\cal A}^{VI}_{1,-2,{\rm nn}} &=& -\alpha \left(E_0^{(1)} -E_0^{(2)}\right)
\left[ \frac{{\cal A}_{-1}}{C_0}\right]^2~, \\
\Delta {\cal A}^{VI}_{1,-2,{\rm pp}} &=& -\alpha \left(E_0^{(1)} +E_0^{(2)}\right)
\left[ \frac{{\cal A}_{-1}}{C_0}\right]^2~, 
\eea
where the coupling constants $E_0^{(1,2)} (\mu )$ obey the renormalization group
equations,
\bea
\mu \, \alpha \,\frac{dE_0^{(1)} (\mu )}{d\mu} 
&=& \alpha \frac{M}{2\pi} E_0^{(1)} (\mu ) 
C_0 (\mu )\mu - \delta m^2
\frac{M g_A^2}{48 \pi^2 f_\pi^2} C_0^2 (\mu)~, \nonumber\\   
\mu \, \alpha \, \frac{dE_0^{(2)} (\mu )}{d\mu} &=& 
\alpha \frac{M}{2\pi} E_0^{(2)} (\mu )  C_0(\mu ) \mu~.
\eea
The relation of the $pp$ and $nn$ scattering lengths to the $np$ one reads
(of course, in the $pp$ system Coulomb subtraction is assumed),
\bea\label{aE}
\frac{1}{a_{pp}} &=&  \frac{1}{a_{np}}  -\frac{4\pi\alpha (E_0^{(1)}
+ E_0^{(2)})}{MC_0^2} + \Delta~,\nonumber\\
\frac{1}{a_{nn}} &=&  \frac{1}{a_{np}}  -\frac{4\pi\alpha (E_0^{(1)}
- E_0^{(2)})}{MC_0^2} + \Delta~,\\
\Delta &=& \delta m^2 \frac{g_A^2 (-C_0 M(m-2\mu ) + 8\pi + C_0 M m \ln
  (m^2/\mu^2))}{12 \pi m C_0 f_\pi^2}~.\nonumber
\eea
For the effective ranges, we have only CIB  
\be\label{RE}
r_{nn} = r_{pp} =  r_{np} + 
\delta m^2 \frac{g_A^2 (C_0 M \mu  +4\pi )( C_0 M (3\mu -2 m) 
+ 12\pi)}{6 \pi M m^4 C_0^2 f_\pi^2}~.
\ee
Note that the relations in eqs.(\ref{aE},\ref{RE}) are  perturbatively
scale--independent and that the last one does not contain
any unknown parameter. We remark that for the CIB scattering lengths difference
the pion contribution alone is {\it not} scale--independent and can thus
never be uniquely disentangled from the contact term contribution 
$\sim E_0^{(1)}$.
While the leading OPE contribution resembles the result obtained in meson 
exchange models, the mandatory appearance of this contact term is a 
distinctively  new feature of the effective field theory approach. 
Matters are somewhat different for CSB. Here, to leading order there is 
simply a four--nucleon contact term proportional to the constant $E_0^{(2)}$.
Its value can be determined from a fit to the empirical
value given in eq.(\ref{CSBval}). Corrections to the leading
order CIB and CSB effect are classified below. We add the following remarks.
Eqs.(\ref{aE}) and (\ref{RE}) are not exact but hold perturbatively. Similarly,
the phase shifts are also defined perturbatively and therefore
one can not have an exact effective range expansion of $\delta = \delta_0 + \delta_1$
(referring to the isospin symmetric case for the moment). If one works e.g. at
next--to--leading order, the norm of the S--matrix is not exactly one,
$S^\dagger S = 1 + {\cal O}(\tilde{Q}^2) \ne 1$. From this it follows that
if one adjusts the parameters so as to reproduce the phase shifts in a certain
momentum range, one can only approximately reproduce the scattering length 
and the effective range, $a \simeq a_{\rm exp}$ and $r \simeq r_{\rm exp}$,
although  the phase shifts may be very well described. This should be
kept in mind.

We are now in the position to calculate the phase shifts. In total, we have five
coupling constants, three of them related to the isospin--symmetric
case, called $C_0$, $C_2 $ and $D_2$. We fit the
paramters $C_{0,2}, D_2$ to the $np$ phase shift under the condition
that the scattering length and effective range are exactly reproduced.
The two new parameters $E_0^{(1,2)}$ are determined from the $nn$ and
$pp$ scattering lengths. The resulting parameters at $\mu = m$ are of 
natural size,
\bea\label{Cval}
C_0 &=& -3.46~{\rm fm}^2~, \quad
C_2 =  2.75~{\rm fm}^4~, \quad
D_2 =  0.07~{\rm fm}^4~, \nonumber \\
E_0^{(1)} &=&  -6.47~{\rm fm}^2~, \quad
E_0^{(2)} =   1.10~{\rm fm}^2~.
\eea
The $np$ and $nn$ $^1S_0$ phase shifts are shown in fig.~2.
To arrive at these curve, the physical masses for the proton and the 
neutron have been used. We stress
again that the effect of the em terms $\sim E_0^{(1,2)}$ is small
because of the explicit factor of $\alpha$ not shown in eq.(\ref{Cval}). 
Having fixed these parameters, we can now predict the $nn$ $^1S_0$ phase 
shift as depicted by the solid line in fig.2. It agress with $nn$
phase shift extracted from the Argonne V18 potential (with the
scattering length and effective range exactly reproduced) up to
momemta of about 100~MeV. The analogous curve for
the $pp$ system (not shown in the figure) is close to the solid line
since the CSB effects are very small.
\begin{center}
\begin{figure}[htb]
\centerline{\psfig{file=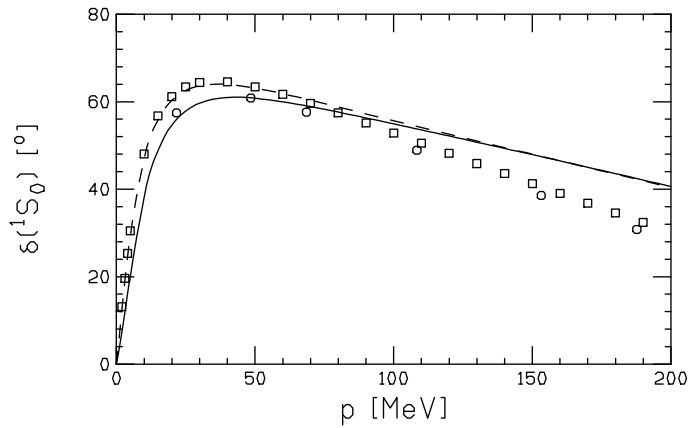,height=2.2in}}
\caption{
$^1S_0$ phase shifts for the $np$ (dashed line) and $nn$ (solid
line) systems versus the nucleon cms momentum. 
The emipirical values for the $np$ case (open squares)
are taken from the Nijmegen analysis. The open octagons
are the $nn$ ``data'' based on the Argonne V18 potential.}
\end{figure}
\end{center}

\section{Classification scheme}

In the framework presented here, it is straightforward to work
out  the various leading (LO), next--to--leading
(NLO) and next--to--next--to--leading order (NNLO) contributions to
CIB and CSB, with respect to the expansion in $\tilde{Q}$, to leading order
in $\alpha$ and the light quark mass difference. Here, I will simply
enumerate the pertinent contributions which appear at a given order.
This naturally also gives an estimate about their relative numerical
importance based on simple dimensional analysis. Based on such an analysis,
one can shed some light on the puzzles discussed in section~2, in particular
the size of the $\pi\gamma$ graphs and the role of charge--dependent
$\pi N$ coupling constants. Let me first consider CIB.
In the classification scheme given below, TPE/3PE stands for 
two/three--pion--exchange.

\medskip

\begin{center}
\begin{tabular}{|l|l|}\hline
LO $\quad\qquad$& Pion mass difference in OPE, 
\\ $\alpha \tilde{Q}^{-2}$ 
& four--nucleon contact interaction with no derivatives.\\ 
\hline
NLO & Pion mass difference in TPE, 
\\ $\alpha \tilde{Q}^{-1}$, $\varepsilon  \tilde{Q}^{-1}$ 
& four--nucleon contact interaction with two derivatives,\\
& insertions proportional to the $np$ mass difference,\\ 
\hline
NNLO & Pion mass difference in 3PE, 
\\ $\alpha \tilde{Q}^{0}$ 
& four--nucleon contact interaction with four derivatives,\\ 
& CIB in the pion--nucleon coupling constants,\\ 
& $\pi \gamma$--exchange.   \\ \hline
\end{tabular}
\end{center}

\medskip

\noindent
Note that there are no CIB effects due to the light quark mass difference
linear in $\varepsilon = m_u - m_d$ from OPE and from the
four--nucleon contact terms. These type of terms can only appear to
second order in $\varepsilon$. 
Effects due to the charge dependence of the pion--nucleon coupling constants,
i.e. isospin breaking terms from ${\cal L}_{\pi N}^{\rm em}$, only start to
contribute at order $\alpha \tilde{Q}^0$. Such effects are therefore 
suppressed by two orders of $\tilde{Q}$ compared to the leading terms,
i.e. by a numerical factor of about $(1/3)^2$. This finding is in agreement
with the various numerical  analyses performed in potential models (if
one assumes that there is indeed some charge dependence in the pion--nucleon
coupling constants, see e.g. the discussion in ref.\cite{jerry}.) 
We remark that the TPE contribution is nominally suppressed by one
order in the small parameter, i.e. one would expect its contribution to
be one third of the leading OPE with different pion masses. This agrees with 
some (but not all) meson--exchange model calculations.
\begin{center}
\begin{figure}[htb]
\centerline{\psfig{file=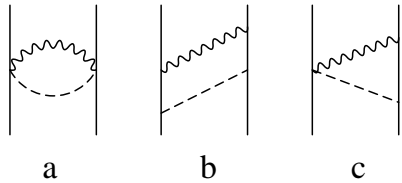,height=1truein}}
\caption{
$\pi \gamma$ graphs. For CIB, the leading contribution comes from the
three diagrams a), b) and c). Due to the  Kroll--Ruderman vertex in
a) and c) the exchanged pion is charged. In case of CSB, the dominant
contribution comes from b) with neutral pion exchange. c) vanishes due
to parity and a) is suppressed by  two additional powers in
$\tilde{Q}$. Crossed graphs are not shown. Also omitted
are the initial and final--state interactions between the nucleon.}
\end{figure}
\vspace{-0.3cm}
\end{center}
Some remarks on the $\gamma \pi$ graphs (some of them shown in fig.~3) are
in order. First, we note that their contribution for CIB is suppressed
by two powers in $\tilde{Q}$ with respect to the leading 
terms (the power counting for these graphs is detailed in appendix~B). The
larger contribution found in some meson--exchange model calculations (approximately
$1/3^{\rm rd}$ of the leading OPE contribution) is therefore at first sight 
at variance with dimensional analysis since one would expect a
suppression of about $(1/3)^2$.\cite{jerry} However, the much smaller value 
obtained in the more recent calculation of ref.\cite{CF} (performed in the static
limit for the nucleons and Coulomb gauge for the photons), 
$\Delta a_{\rm CIB}(\pi\gamma)/ \Delta a_{\rm CIB}({\rm OPE}) \simeq -1/14.6$,  
is in much better agreement with the suppression by two powers in $\tilde{Q}$. 
This deserves further study.
Second, the $\gamma \pi$ graphs evaluated in heavy baryon chiral
perturbation  theory have been used in a refined potential model
calculation.\cite{birar} The $^1$S$_0$ $np$ low--energy parameters were found to
be very little affected. It is also instructive to see how the
corrections due to the $np$ mass difference come about. For that, let
us write the nucleon mass as $M+ \delta M$, where $\delta M$ subsumes
the em and strong contribution to the $np$ mass splitting, i.e. these
terms are ${\cal O}(\alpha )$ and ${\cal O}(\varepsilon )$. 
Differentiating the leading isospin--symmertic amplitude eq.(\ref{Amo}) with
respect to the nucleon mass shift gives 
\be
{\partial {\cal A}_{-1} \over \partial M} = -{{\cal A}_{-1}^2 \over 4
  \pi} \, (\mu + ip )~,
\ee
from which it follows that
\be
\delta {\cal A}_{-1} = -\delta M {{\cal A}_{-1}^2 \over 4 \pi} \, (\mu
+ ip ) \sim \varepsilon \tilde{Q}^{-1}, \alpha\tilde{Q}^{-1}~.
\ee
We now turn to CSB. The pattern of the various contributions looks different:

\medskip

\begin{center}
\begin{tabular}{|l|l|}\hline
LO & Electromagnetic four--nucleon 
contact interactions \\ $\alpha \tilde{Q}^{-2}, \varepsilon Q^{-2}$ 
& with no derivatives.\\ \hline
NLO &
Em four--nucleon contact terms with two derivatives, \\
$\alpha \tilde{Q}^{-1}, \varepsilon Q^{-1}$ &
 strong isospin--breaking
contact terms w/ two derivatives, \\ & insertions proportional to the
$np$ mass difference.\\ \hline
\end{tabular}
\end{center}

\medskip
\noindent
We remark that the leading order CSB effects do not modify the effective
range but the corresponding scattering length. Also, the leading $\pi \gamma$
graphs (with the exchanged pion now being neutral) are of order
$\alpha \tilde{Q}^{0}$, i.e. suppressed by two orders in the KSW
expansion parameter. This can be used as an argument to simply ignore
such contributions in the study of CSB. Within meson--exchange models,
their contribution comes out small but there seems to be no consensus
about the actual size.\cite{jerry} A similar hierarchy can
also be derived based on the Weinberg power counting (which applies
to the potential) as discussed in ref.\cite{bira}.

\section{The pionless theory}

It is also of interest to consider a simplified theory in which even the
pions are integrated out and one only deals with four--nucleon contact
terms. Such an approximation is clearly only justifiable in a
KSW--like approach as followed here. We do not want to enter the
discussion about how to treat the pions (perturbative versus
nonperturbative) but rather wish to see how accurate the pionless
theory works for the question of isospin violation. In such an
approach the connection to the chiral aspects of QCD is, of course, lost.
Integrating out
the pions modifies the coupling constants of the various contact
terms like

\medskip

\begin{center}
\begin{tabular}{llll}
EFT with pions: & $C_0$, $C_2$, $D_2$ &and& $E_0^{(1)}$, $E_0^{(2)}$~,\\
EFT without pions: & $\tilde{C}_0$, $\tilde{C}_2$ 
&and& $\tilde{E}_0^{(1)}$, $\tilde{E}_0^{(2)}$~.\\
\end{tabular}
\end{center}

\medskip
\noindent
In the pionless theory, there is no more $D_2$ term. One has thus
four parameters which can be fixed from the three scattering lengths
and the $np$ effective range. The values of the parameters are
\bea\label{Cvalnop}
\tilde{C}_0 &=& -3.42~{\rm fm}^2~, \quad
\tilde{C}_2 =  6.05~{\rm fm}^4~,  \nonumber\\
\tilde{E}_0^{(1)} &=&  8.09~{\rm fm}^2~, \quad
\tilde{E}_0^{(2)} =   1.07~{\rm fm}^2~.
\eea
\begin{center}
\begin{figure}[htb]
\centerline{\psfig{file=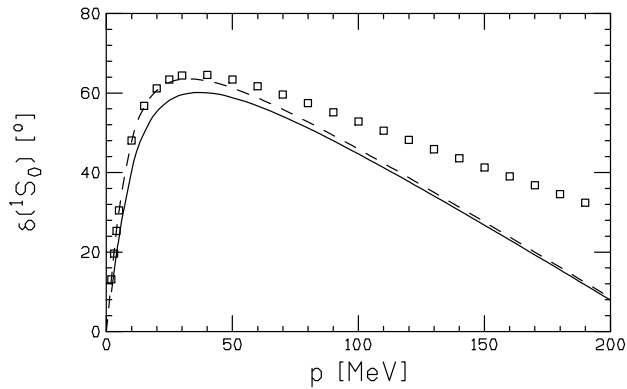,height=2.0in}}
\caption{
$^1S_0$ phase shifts for the $np$ (dashed line) and $nn$ (solid
line) systems versus the nucleon cms momentum in the pionless
theory. The emipirical values for the $np$ case (open squares)
are taken from the Nijmegen analysis.}
\end{figure}
\end{center}
Compared to the values for the theory with pions given in
eq.(\ref{Cval}), we note that the value of $\tilde{C}_0$ almost the
same as the one of $C_0$. This is expected since we are close to a fix
point. Noticeably, there is a dramatic difference in
$\tilde{E}_0^{(1)}$ compared to $E_0^{(1)}$, which is also expected
since the former pion contribution is now completely absorbed in the
contact term. The resulting $np$ and $nn$ $^1S_0$ phase shifts shown 
in fig.~4. The description of the phase shifts has clearly worsened,
they start to deviate from the data for momenta larger than 50~MeV
and are considerably off the data for higher momenta, say about
200~MeV. These results are not surprising, but it is comforting to see
that the pionless theory indeed describes the small momentum region
correctly. For example, to this order the $nn$ effective range equals
the one of the $np$ system which is in agreement with the data
(partial wave analysis). However, the interpretation of the CIB is now very
different since it is entirely given by the isospin--violating 
contact interaction. This makes a comparison with the results obtained
in potential models very difficult, if not impossible.

\section{Summary}
We have shown that isospin violation can be systematically included
in the effective field theory approach to the two--nucleon system in
the KSW formulation. For that, one has to construct the most general
effective Lagrangian containing virtual photons and extend the
power counting accordingly. We have shown that
this framework allows one to systematically classify the various
contributions to charge indepedence and charge symmetry breaking (CIB
and CSB). In
particular, the power counting combined with dimensional analysis 
allows one to understand the suppression of contributions
from a possible charge--dependence in the pion--nucleon coupling constants.
Including the pions, the leading CIB breaking effects are the
pion mass difference in OPE together with a four--nucleon contact term.
These effects scale as $\alpha \tilde{Q}^{-2}$. Power counting lets one
expect that the much debated contributions from two--pion exchange and
$\pi\gamma$ graphs are suppressed by factors of $1/3$ and $(1/3)^2$,
respectively. This is in agreement with some, but not all, previous
more model--dependent calculations. The leading charge
symmetry breaking is simply given by a four--nucleon contact term. One can
also use the pionless theory to describe the observed CIB and CSB,
but the physical interpretation is much less clear. It would certainly
be interesting to investigate such effects in the presence of external
sources.

\section*{Acknowledgments}
I thank the organizers for  providing a stimulating atmosphere and
Jerry Miller for some useful comments.

\vfill


\section*{Appendix~A: The average pion mass}

Here, I briefly show how the average pion mass used in the $np$ system
in section~4 comes about.
Consider the graphs shown in fig.~5. The strong interaction potential
for OPE in the $nn$ and $pp$ system is given by the standard Yukawa
form
\be
V_{nn} = V_{pp} = {g_{\pi N}^2\over 4\pi} {(\vec{\sigma}_{(1)} \cdot \vec{\nabla})
(\vec{\sigma}_{(2)} \cdot \vec{\nabla}) \over 4M^2} {{\rm e}^{-m_0 r} \over r}~,
\ee
where $m_0$ is a shorthand notation for the neutral pion mass. 
\begin{center}
\begin{figure}[h]
\centerline{\psfig{file=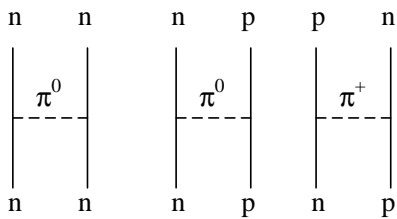,height=1.2in}}
\caption{
OPE graphs for the $nn$ (left) and the $np$ system (middle and right).}
\end{figure}
\vspace{-0.3cm}
\end{center}
\noindent
For  the  $np$ case, we also have  charged pion  exchange,
\be
V_{np} = V_{nn} \left[ -1 + 2{\rm e}^{-\delta m\, r} \right] 
\simeq V_{nn} \left[ 1 - 2 \delta m \, r \right]~,
\ee
with $\delta m = m_\pm - m_0$ and $m_\pm$ denotes mass of the charged pion.
So effectively one has for the  $np$ OPE a mass 
\be
m = m_0 + 2\delta m \quad {\rm or} \quad m^2 = m_0^2 + 2\delta m^2~,
\ee
since $m^2 \simeq m_0^2 + 4\delta m \, m_0 = m_0^2 +4(m_\pm -m_0) m_0 
\simeq m_0^2 + 2\delta m^2$ (note that $\delta m^2 \neq (\delta m)^2$).
These are exactly the formulae used in section~4.

\section*{Appendix~B: Power counting for the $\pi\gamma$ graphs}

In this appendix, we will work out the explicit power counting
of the various $\pi\gamma$--exchange diagrams (cf. fig.~3). First
we need the counting rules for the propagators, vertices etc.:
Pion propagator~$\sim \tilde{Q}^{-2}$, 
nucleon propagator~$\sim\tilde{Q}^{-1}$,
pion-nucleon vertex~$\sim\tilde{Q}$,
photon-nucleon vertex~$\sim \tilde{Q}^0$,
charged~pion-photon-nucleon vertex~$\sim \tilde{Q}^0$,
neutral~pion-photon-nucleon vertex~$\sim \tilde{Q}$ and
loop integration~$\sim \tilde{Q}^{4}$. This gives (displayed
is the respective power in $\tilde{Q}$ for all vertices, propagators,
etc for the specified diagram):

\medskip

\begin{center}
\begin{tabular}{lcccc}
Graph & Vertices & Propagators & Loops & Total\\ \hline 
3a ($\pi^\pm$)& 0 & -4 & 4 & 0 \\
3a ($\pi^0$) & 2 & -4 & 4 & 2 \\
3b ($\pi^\pm$) & 2 & -6 & 4 & 0 \\
3b ($\pi^0$) & 2 & -6 & 4 & 0 \\
3c ($\pi^\pm$) & 1 & -5 & 4 & 0 \\
3c ($\pi^0$) & 2 & -5 & 4 & 1 \\
\end{tabular}
\end{center}

\medskip

\noindent
For CIB, we have charged and neutral pion exchange and all $\pi\gamma$
diagrams 3a,b,c start  contributing $\tilde{Q}^0$. In the case of CSB,
matters are somewhat different. Graph 3b is leading at $\tilde{Q}^0$,
while 3c is formally of order $\tilde{Q}$ but vanishes due to parity.
The contribution of graph~3a to CSB is suppressed by two additional
powers of the small momentum.


\end{document}